\documentclass[aps,pre,twocolumn, superscriptaddress,showpacs,amsmath,amssymb]{revtex4}

\usepackage{graphicx}
\usepackage{dcolumn}
\usepackage{bm}

\usepackage{epstopdf}

\usepackage{color}

\usepackage{bm}
\usepackage[normalem]{ulem}

\usepackage[space]{grffile}

\newcommand{\ii}{{\rm i}}

\newcommand{\bu}{\mathbf{u}}

\newcommand{\sep}{ \ \ \ , \ \ \ }

\newcommand{\beq}{\begin{equation}}
\newcommand{\eeq}{\end{equation}}
\newcommand{\beqn}{\begin{eqnarray}}
\newcommand{\eeqn}{\end{eqnarray}}

\newcommand{\ee}{{\rm e}}
\newcommand{\eq}{Eq.~}

\newcommand{\fig}{Fig.\ }

\begin{document}

\title{
	Edge instability in incompressible planar active fluids}

\author{David Nesbitt}
\affiliation{Department of Bioengineering, Imperial College London, South Kensington Campus, London SW7 2AZ, U.K.}
\author{Gunnar Pruessner}
\affiliation{Department of Mathematics,  Imperial College London, South Kensington Campus, London SW7 2AZ, U.K.}
\author{Chiu Fan Lee}\email{c.lee@imperial.ac.uk}
\affiliation{Department of Bioengineering, Imperial College London, South Kensington Campus, London SW7 2AZ, U.K.}

\date{\today}

\begin{abstract}

Interfacial instability is highly relevant to many important biological processes. A key example arises in wound healing experiments, which observe that an epithelial layer with an initially straight edge does not heal uniformly. We consider the phenomenon in the context of active fluids. 
Improving upon the approximation used in  J.~Zimmermann, M.~Basan and H.~Levine, {\it Euro. Phys. J.: Special Topics} {\bf 223}, 1259 (2014), we perform a linear stability analysis on a two dimensional incompressible hydrodynamic model of an active fluid with an open interface. We categorise the stability of the model and find that for experimentally relevant parameters, fingering stability is always absent in this minimal model. Our results point to the crucial importance of density variation in the fingering instability in tissue regeneration. 

%
\end{abstract}

\maketitle

The Saffman-Taylor instability is a classic example of interfacial pattern formation in fluid dynamics. Also known as viscous fingering, it refers to the interfacial phenomenon  
observed when a fluid is injected into a Hele-Shaw cell, displacing a resting fluid of higher viscosity. The 
leading edge of the advancing fluid does not propagate uniformly, but splits into fingerlike protrusions
 \cite{Saffman1958}. Whilst this is well understood in the context of classical fluids, what happens when we instead use an active fluid?

Active matter refers to any system of interacting particles in which one or more of the agents present can expend stored or ambient free energy to generate some form of self-propulsion \cite{Marchetti2013}. The most obvious examples of these systems are found in biology and span many length scales, such as flocks of birds \cite{Toner1995}, bacterial suspensions \cite{Hatwalne2004} and the cytoskeleton of a eukaryotic cell \cite{Julicher2007}. In particular, epithelial tissue can be regarded as active matter \cite{Zimmermann2014}. In its simplest form, as in wound healing experiments, the tissue consists of a single layer of tightly packed cells. These cells not only interact through adhesion and contact forces, they can also generate motility forces by crawling on the substrate.

\begin{figure}[h]
 	\centering
   \includegraphics[width=0.35\textwidth]{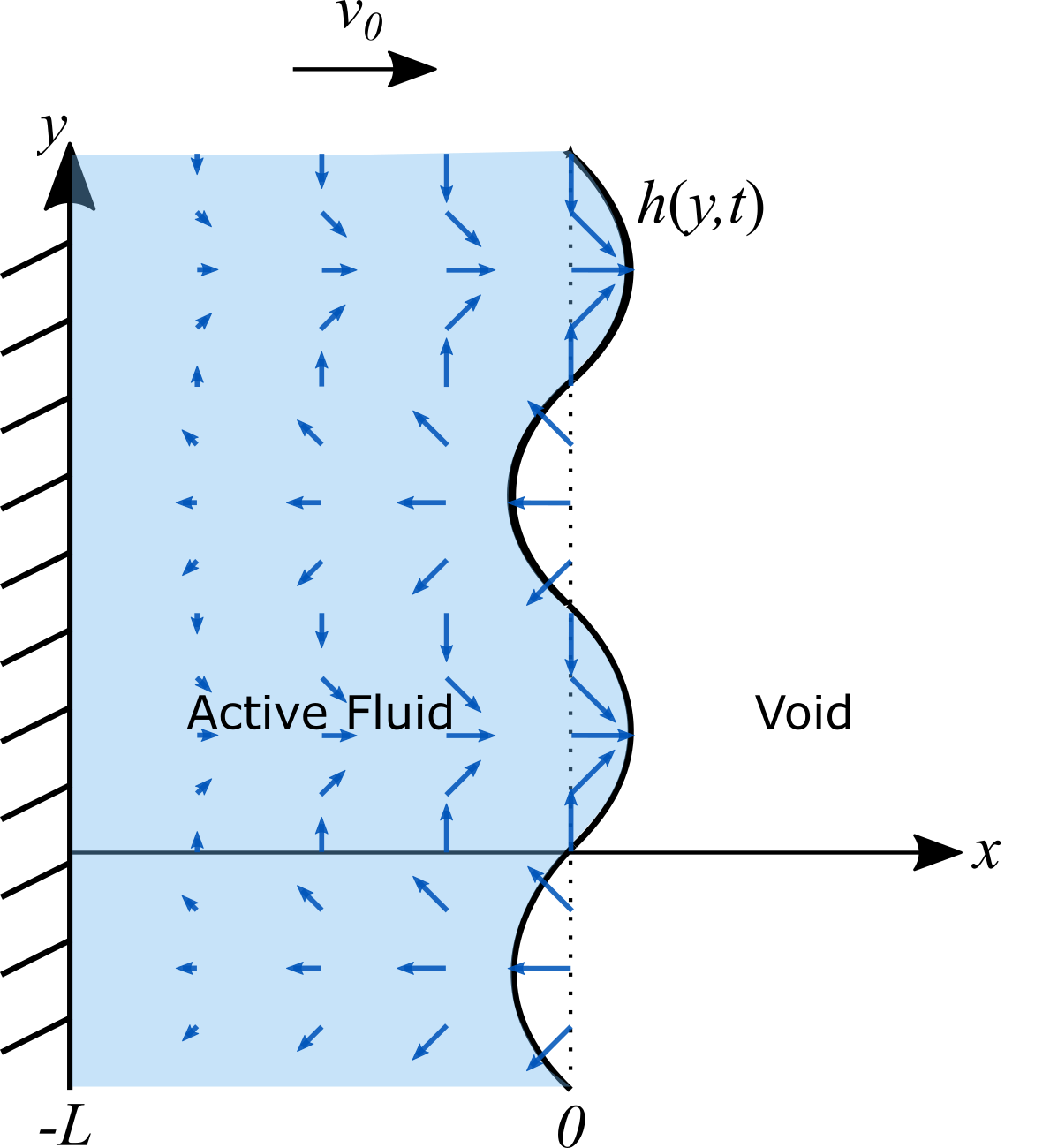} 
  \caption{   Schematic of the model geometry. A two dimensional strip, with thickness $L$, of active fluid, bounded between a solid wall and an empty void, is permitted to move with speed $v_0$ in the direction of the void. The trailing wall comoves with the bulk fluid. }
  \label{fig:ModelDiagram}
  \end{figure}

The study of edge stability of an active fluid is thus highly relevant to many important biological processes. For instance, when  an {\it in vitro} cell monolayer  is scratched, or barricades are removed \cite{Petitjean2010,Poujade2007}, simulating a wound, the tissue spreads to fill the void (\fig \ref{fig:ModelDiagram}). However, a common observation from experiments is that an initially straight wound does not heal uniformly \cite{Petitjean2010,Poujade2007}. Instead finger-like protrusions develop at the leading edge, reminiscent of those in the Saffman-Taylor instability. The exact reason for this pattern formation remains unclear, with some attributing the effect to particular `leader' cells guiding the rest \cite{Gov2007,Poujade2007,Omelchenko2003}. However, simulations have shown that the phenomenon can emerge from the collective motion of actively crawling cells with strong cell-cell adhesion, with no need for leader cells \cite{Basan2013}. Other studies have investigated the interface between competing tissues; their simulations suggest that the interface advances at a constant speed and does not appear to form fingers, merely fluctuating within a stable region of roughness \cite{Podewitz2016}.
In particular, a recent linear stability analysis of a hydrodynamic model suggested that the edge of a homogeneous incompressible planar active fluid is stable when the fluid is moving at a constant rate, however the interface becomes unstable if the fluid is initially stationary \cite{Zimmermann2014}. Here, we perform a more thorough theoretical analysis of the same model as in \cite {Zimmermann2014} and arrive at a qualitatively different conclusion.

The model system consists of an incompressible two-dimensional active fluid propagating in the direction of its free surface, as illustrated in \fig \ref{fig:ModelDiagram}. This setup is similar to that of Saffman-Taylor, however we do not apply a pressure gradient to the active fluid; any propulsion will be self generated. We also assume that the fluid being displaced is of negligible viscosity and density. The flow in this region is therefore not solved for and instead is treated as a space of constant pressure. 

The flow field, $\bu$, of the active fluid is described  by the deterministic, incompressible version of the Toner-Tu equations \cite{Toner1995,Toner1998,Toner2005}
    \begin{subequations} \label{eqn:TonerTu}
   		\begin{align}
   \rho \left( {\frac{{\partial \mathbf{u}}}{{\partial t}} + {\lambda}\left( {\mathbf{u} \cdot \nabla } \right)\mathbf{u}} \right) &= \mu {\nabla ^2}\mathbf{u} - \nabla p + a \mathbf{u} - b {\left| \mathbf{u} \right|^2}\mathbf{u} \ , \label{eqn:TonerTu1} \\
     \nabla  \cdot \mathbf{u} &= 0 \ ,\label{eqn:TonerTu2} 
  		 \end{align}
   \end{subequations}
where $\rho, \mu$ are the density and viscosity of the active fluid respectively; $p$ is the pressure, treated here as a Lagrange multiplier to enforce the incompressbility condition \eqref{eqn:TonerTu2}. The two terms $a \mathbf{u}$ and $b {\left| \mathbf{u} \right|^2}\mathbf{u}$ account for the activity, where $a$ and $b$ are constants. Each of these acts in a direction tangential to the instantaneous velocity of the fluid. 
The former acts as a driving term with $a>0$ inherently assuming that the propulsion forces within the fluid align with the instantaneous velocity; $a<0$ means that the active forces are acting against the motion.
The non-linear term $b {\left| \mathbf{u} \right|^2}\mathbf{u}$ provides resistance, preventing arbitrary growth, hence $b>0$ necessarily.
 Unlike conventional fluid mechanics, this system does not conserve momentum. As a result, Galilean invariance does not apply and the prefactor of the advective term, $\lambda$, need not be unity. Mass is conserved in the active fluid via the incompressibility condition \eq \eqref{eqn:TonerTu2} and the density $\rho$ is assumed to be constant - a fair assumption as cell division and death rates are low in comparison with the effects of motility forces in wound healing assays \cite{Poujade2007,Zimmermann2014}. Eq.~\eqref{eqn:TonerTu} is rich in physics: one of us has previously shown that the fluctuating form of the equation is connected to the Kardar-Parisi-Zhang model in 2D \cite{chen16} and its associated critical behaviour is described by a novel universality class in non-equilibrium physics \cite{chen15}. Here, we will focus on the stability criteria when an interface is present.

We now perform a linear stability analysis on these equations using the geometry depicted in \fig \ref{fig:ModelDiagram}: a strip of active fluid of thickness $L$, bounded by a solid wall on one side with an open interface on the other. The fluid is assumed to move with a constant uniform base flow $v_0$ in the direction of the open interface, with the rear wall co-moving with the fluid. In such a homogeneous flow, $\mathbf{u}  = v_0\hat{x}$, all of the derivatives in \eq \eqref{eqn:TonerTu} vanish, leaving a balance between the active terms. The solution $v_0=0$ exists for all real $a$, however $a>0$ has the additional solution: $v_0 = \sqrt{a/b}$. We will refer to the $v_0=0$ case as the {\it stationary case} and the $v_0>0$ as the {\it moving case} respectively.

We add a small perturbation to this flow
$\mathbf{u}  = (v_0 + u_x)\hat{x} + u_y \hat{y} $
and the interface 
$h = v_0 t + \tilde{h}$, 
where $|\tilde{h}|,|u_x|,|u_y| \ll 1.$
We consider sinusoidal perturbations with wavenumber $q$, using the form 
\begin{subequations} \label{eqn:linstab}
\begin{align}
u_x &= A \ee^{ r ({x} - v_0 t) + \omega {t} + \ii q {y}} \\
u_y &= B \ee^{ r ({x} - v_0 t) + \omega {t} + \ii q {y}} \\
p &= C \ee^{ r ({x} - v_0 t) + \omega {t} + \ii q {y}} \\
\tilde{h} &= h_0 \ee^{ \omega {t} + \ii q {y}} 
\end{align}
\end{subequations}
so that the real part of $\omega$ describes the growth rate of the mode, with a positive value corresponding to instability.

Substituting \eq (\ref{eqn:linstab}) into \eq \eqref{eqn:TonerTu} results in a homogeneous system of linearised equations in the constants $A,B$ and $C$. In order to have a non-trivial solution the determinant of their matrix of coefficients must be zero. This restricts the values of $r$ to be the roots of the quartic polynomial
\beqn \label{eqn:Quartic}
0&=& \mu {r^4} + \rho \left( 1 - \lambda\right) {v}_0 r^3  
\\
&& - \left[ \rho \omega + { 2 \mu q^2} - \left( a - b {v}_0^{2} \right) \right] r^2 
 - \rho \left( 1 - \lambda\right) {v}_0 q^2 r 
 \nonumber
 \\
 &&+ q^2 \left[ \mu {q^2} + \rho \omega - \left( a - b {v}_0^{2} \right) + 2 b {v}_0^{2} \right]  \ . \nonumber
\eeqn
Hence the general solution for the velocity perturbation is
\beq \label{eqn:GenSolForm}
u_x = \sum_{j=1}^{4} A_j \ee^{r_j ({x} - v_0 t) + \omega {t} + \ii q {y}}
\eeq
where $r_j$ are the four roots in \eq \eqref{eqn:Quartic}. The other perturbation fields $u_y$ and $p$ can also be expressed in terms of $A_j$ using \eq \eqref{eqn:TonerTu}.

The remaining unknowns $A_j$ and $h_0$ are fixed by the boundary conditions. No-slip is applied at the rear wall, meaning all velocity perturbations $(u_x,u_y)$ in the flow field must decay this far from the interface and thus at the rear wall $x = v_0t - L$	
\begin{equation}
{u_x\Big|_{x = v_0t - L}=u_y\Big|_{x = v_0t - L}=0}
\ .
\end{equation}
The velocity of the interface $h(y,t)$ must be continuous with the flow field. Linearised at the free surface equilibrium, $x = v_0t$, this becomes
\beq \label{eqn:BCInterfaceVel} 
\frac{\partial {\tilde{h}}}{\partial {{t}}} = {u_x\Big|_{x = v_0t}} 
\ .
\eeq
The fluid occupying the void region in \fig \ref{fig:ModelDiagram} is of negligible viscosity compared to that of the active fluid. Hence this region is not solved for and is assumed to be of constant pressure. Therefore the tangential stress on the interface must be zero, \eq \eqref{eqn:BCTransverseStress}. The normal stress must be balanced by the surface tension $\gamma$ and the pressure difference across the interface, \eq \eqref{eqn:BCSurfaceTension}. Linearised about $x = v_0t$ these conditions are
\begin{subequations} \label{eqn:BoundaryConditions}
\beqn
\left.  \frac{\partial {u_x}}{\partial {{y}}}\right|_{x = v_0t} + \left.\frac{\partial {u_y}}{\partial {{x}}}\right|_{x = v_0t} &=& 0 \ ,  \label{eqn:BCTransverseStress}   \\
\left. 2 \mu \frac{\partial {u_x}}{\partial {{x}}}\right|_{x = v_0t} - \left. {p}\right|_{x = v_0t} &=& \gamma   \frac{\partial^{2} {\tilde{h}}}{\partial {\hat{y}}^{2}}
\ . \label{eqn:BCSurfaceTension}
\eeqn
\end{subequations}

The five boundary conditions give another system of homogeneous linear equations, this time in terms of the four $A_j$ and $h_0$ (see SM I). Again, for a non-trivial solution the matrix of coefficients, denoted by $M$ and defined in \eq (S2), must have zero determinant. We now let $f(\omega,q)=\det{M}$ and seek to find roots of $f(\omega,q)$.

{\it Stationary case.}~~In the stationary case, $v_0=0$, factoring out non-zero constants reduces $f(\omega,q)=0$ to 
\begin{align} \label{eqn:StationaryFull}
& 0 =  \frac{\mu \omega}{\gamma}   \Bigg\{   - k \cosh (L(k-q)) \frac{\rho^2}{\mu^2}\left( \omega - \frac{a}{\rho} \right)^2    \\
& + k \Big( 1 -  \cosh (L(k-q)) \Big) \left[8q^4 + 4q^2 \frac{\rho}{\mu}\left( \omega - \frac{a}{\rho} \right)\right] \nonumber	 \\
&	 - \sinh (Lk) \sinh (Lq)  (k-q)^2 \left(k^3 + k^2 q + 3 k q^2 - q^3\right) \Bigg\} \nonumber \\
&  -  \frac{\rho}{\mu}\left( \omega - \frac{a}{\rho}  \right)q^3  \Big\{k \cosh (Lk) \sinh (Lq)  \nonumber \\
& \qquad \qquad \qquad - q \sinh (Lk) \cosh (Lq) \Big\} \nonumber
\end{align}
where
\beq \label{eqn:k}
k  = \sqrt  {{q^2} + \frac{\rho}{\mu} \left( \omega - \frac{a}{\rho}  \right)}.
\eeq
We first consider the long wavelength limit (small $q$). Expanding the terms in powers of $q$ we obtain an asymptotic expansion for $\omega$: 
%
\beqn \label{eqn:Expansion}	
\nonumber
		\omega &=& \frac{a}{\rho} - \frac{\mu}{\rho L^2} \left(n \pi + \frac{\pi}{2}\right)^2  \\
		&&
+		\frac{\mu}{\rho}\left(1 + \frac{16 (-1)^{n+1}}{(1+2n)\pi}\right) q^2 + O(q^4)
\eeqn
where $n$ is any integer. It follows that $\omega$ is real for sufficiently small $q$. Our numerical calculations suggest that this holds for all real $q$ and we will assume that $\omega$ is real henceforth.
 
With this assumption we can show by contradiction that $\omega$ is bounded above by $a/\rho$. If we assume that $\omega$ is greater than $  a/\rho$, \eq (\ref{eqn:k}) implies that $k>q$. Upon inspection we find that each line of \eq \eqref{eqn:StationaryFull} is strictly negative, and thus the right hand side cannot be zero. This means there are no solutions if $\omega>a/\rho$.  This in particular resolves the diverging growth rate encountered in \cite{Zimmermann2014}. Our result also provides a finite range in which to search for unstable modes.
 \begin{figure} 

 	 \centering

 	 \includegraphics[width=8.7cm]{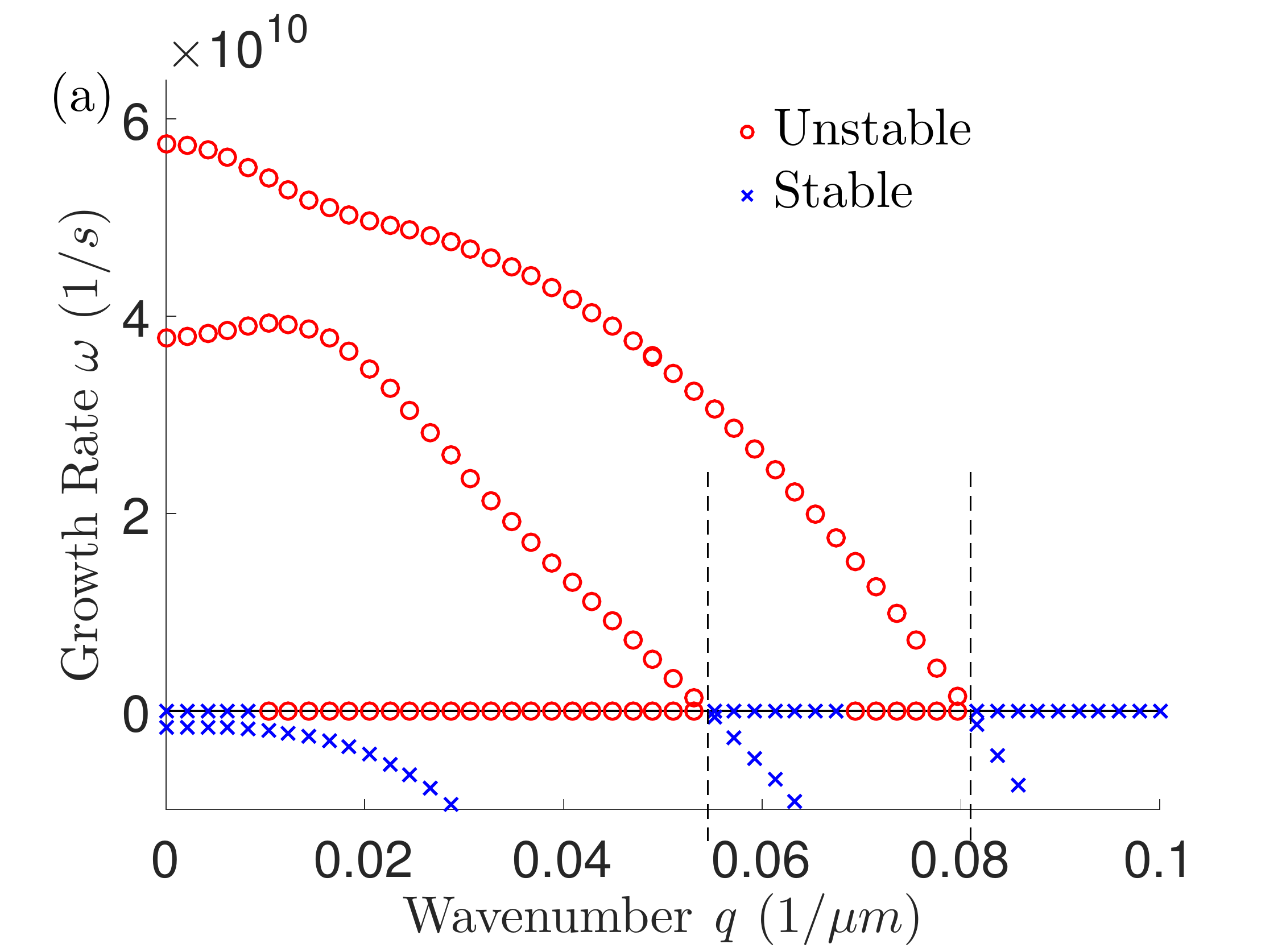} 

 	\includegraphics[width=8.7cm]{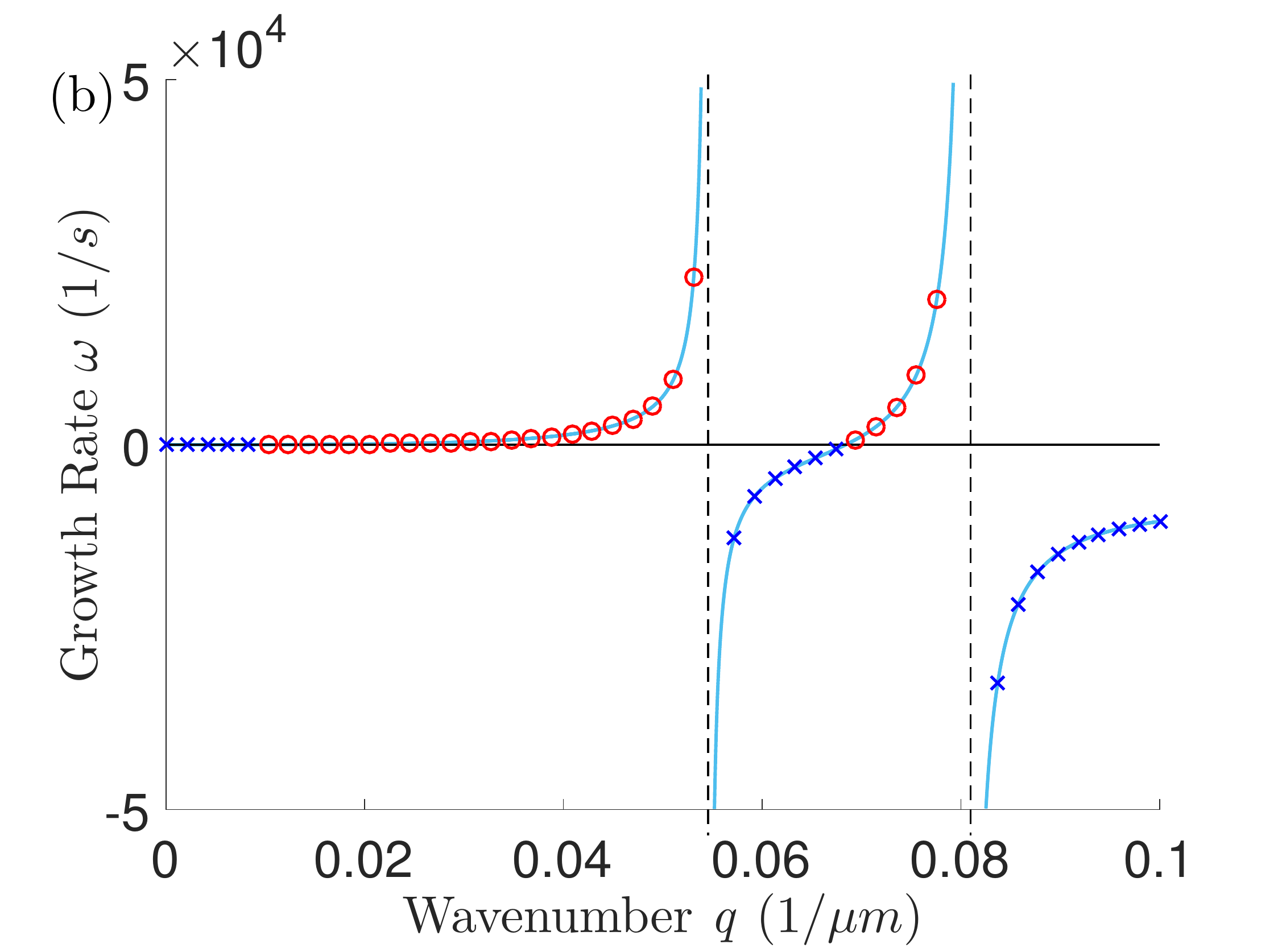}
 	
   \caption{ Stability diagram for the stationary case ($v_0=0$). Red circles indicate unstable solutions, $\omega > 0$; blue crosses indicate stable solutions, $\omega <0$. Plot (b) is the same as (a) but focused on a smaller range of $\omega$. The solid blue curves in (b) show the results obtained when the inertial terms are ignored \cite{Zimmermann2014}. The dashed lines are added as visual aides.
  The physiologically relevant parameter values are: 
    		$\rho = 10^3 \text{kg}\text{m}^{-3}$, $ \mu = 10^4 \text{Pa s}$ \cite{Forgacs1998,Schotz2008}, $ \gamma = 10^3  \text{ Pa }\mu\text{m}$ \cite{Foty1994} and $L=100\mu$m \cite{Petitjean2010}.
    		The steady-state speed of the tissue, $\sqrt{a/b}$, is $2.7 \times 10^{-3}\text{ }\mu\text{m s}^{-1}$ \cite{Poujade2007}. 
    		 $\lambda$ is irrelevant in the stationary case.
    		 For $a$ and $b$, we use the parameters employed in \cite{Zimmermann2014}: $a = 60 \text{ Pa s }\mu\text{m}^{-2}, b = 10^7 \text{ Pa s}^3\mu\text{m}^{-4}$. This choice of $a$ satisfies the condition in \eq \ref{eqn:InstabilityCondition} for instability, and the choice of $b$ sets the steady-state speed to be $2.45 \times 10^{-3} \text{ }\mu\text{m s}^{-1}$, in line with the physiological value above. This also facilitates comparison with the results of \cite{Zimmermann2014}. Indeed, according to \eq \eqref{eqn:InstabilityCondition} the minimum $a$ for instability is about $2.5 \text{ Pa s }\mu\text{m}^{-2}$, and as we will explain in Summary \& Discussion, is likely too large to be physiologically relevant.		
    }
  \label{fig:Stationary}
  \end{figure}

\fig \ref{fig:Stationary}(a) depicts the full unstable solution for the specified parameter values, obtained by solving $f(\omega,q)=0$ numerically. Where necessary, we have excluded the zeros of $f$ corresponding to degenerate cases of \eq \eqref{eqn:Quartic}. These must be treated separately to determine if they are genuine solutions (see SM II). 

As the growth rate is now bounded above by $a/\rho$, there are no singularities as discussed in \cite{Zimmermann2014}, and we seek the wavenumber that provides the largest growth rate. The highest curve outlined in \fig \ref{fig:Stationary}(a) appears to obtain its maximum in the limit as $q 
\to 0$. This is in accord with the expansion in \eq \eqref{eqn:Expansion} -- the constant part of $\omega$ is largest when $n = 0$, at which point the coefficient of the next lowest order term, $q^2$, is negative. As such the growth rate initially decays away from its peak at $q=0$. The actual value $q=0$ is not a valid mode. This would correspond to a flat wave, which would necessarily violate conservation of mass if it were to grow.
 
To understand why the most unstable mode occurs as $q$ goes to 0, recall that there is no external pressure, and the only driving force in our system is the internally generated active force $a \bu$. This force provides the same growth rate for all wavenumbers $q$. The motion is resisted by viscosity and surface tension. The strength of each of these increases with $q$, thus suggesting that the maximum growth should be in their absence, namely in the limit $q \to 0$.

Viscosity counteracts shear within the velocity field. Due to the no-slip condition at the rear wall coupled with the incompressibility condition, any velocity perturbation at the free surface naturally induces shear throughout the fluid, as depicted schematically in \fig \ref{fig:ModelDiagram}. As $q$ drops to zero, the shear due to variation in the $y-$direction becomes negligible, however the shear in the $x-$direction remains, as the velocity must vanish across the width $L$. Hence the very minimum we require for any perturbation to grow, is for the driving force $a \bu$ to be stronger than the viscosity component corresponding to the latter direction, $\mu {\partial^2 {\mathbf{u}}}/{\partial {{x}^2}}$. We therefore expect instability to occur only if
\beq
a U > C \frac{\mu U}{L^2} .
\eeq
for some unknown dimensionless constant $C$.
We can quantify $C$ by returning to the asymptotic expansion \eqref{eqn:Expansion}. Since $\max_q \omega$ is achieved for $n=0$ as $q \rightarrow 0$, we indeed find that the stationary case is linearly unstable only if 
\begin{gather} \label{eqn:InstabilityCondition}
\frac{a L^2}{\mu} > \frac{\pi^2}{4}.
\end{gather}
Hence we have a condition for instability. Evidently a negative value of $a$ implies the system is stable -- an expected result as both active terms would be acting against the motion.

We have concluded that the most unstable mode is simply the largest wavelength that the domain permits; our model assumed the strip to be infinitely long, hence no restriction was placed on the wavenumber. In the context of wound healing, 
our model suggests that the typical thickness for the 
growing fingers will be of the system size 
and thus does not generate the behaviour exhibited in wound healing experiments. 
We will comment on this further in Summary \& Discussion.

Fig. \ref{fig:Stationary}(b) focuses on a region of much smaller $\omega$. Also plotted are the results obtained using the analysis of \cite{Zimmermann2014}. Within this region the solutions actually align very well, revealing that their simplification to ignore the inertial terms in their analysis is valid when $|\omega|$ is `small'. We can clarify the meaning of small by returning to the governing equation \eqref{eqn:TonerTu1}. Deeming $ \rho \partial \mathbf{u}/\partial t $ to be negligible with respect to $ a \mathbf{u}$ is to assume that $|\omega| \ll a/\rho$, hence their solution is valid only within this region.
Ultimately this means that although the flow has a very small Reynolds number Re, it also has a very small Stokes number St, which are given by the ratios
 \beq
\text{Re} = \frac{\rho\left( {\mathbf{u} \cdot \nabla } \right)\mathbf{u}}{ \mu {\nabla ^2}\mathbf{u}} \qquad \text{and} \qquad \text{St} = \frac{ \left( {\mathbf{u} \cdot \nabla } \right)\mathbf{u}}{{{\partial \mathbf{u}}}/{{\partial t}} }   .
  \eeq
 If $\text{St}\sim \mathcal{O}(1)$ and $\text{Re}\ll 1$, both of the terms ${ \left( {\mathbf{u} \cdot \nabla } \right)\mathbf{u}}$ and ${{{\partial \mathbf{u}}}/{{\partial t}} }$ would be negligible, however in this case only the former is.

 \begin{figure}[h!]
 	\centering

 \includegraphics[width=8.7cm]{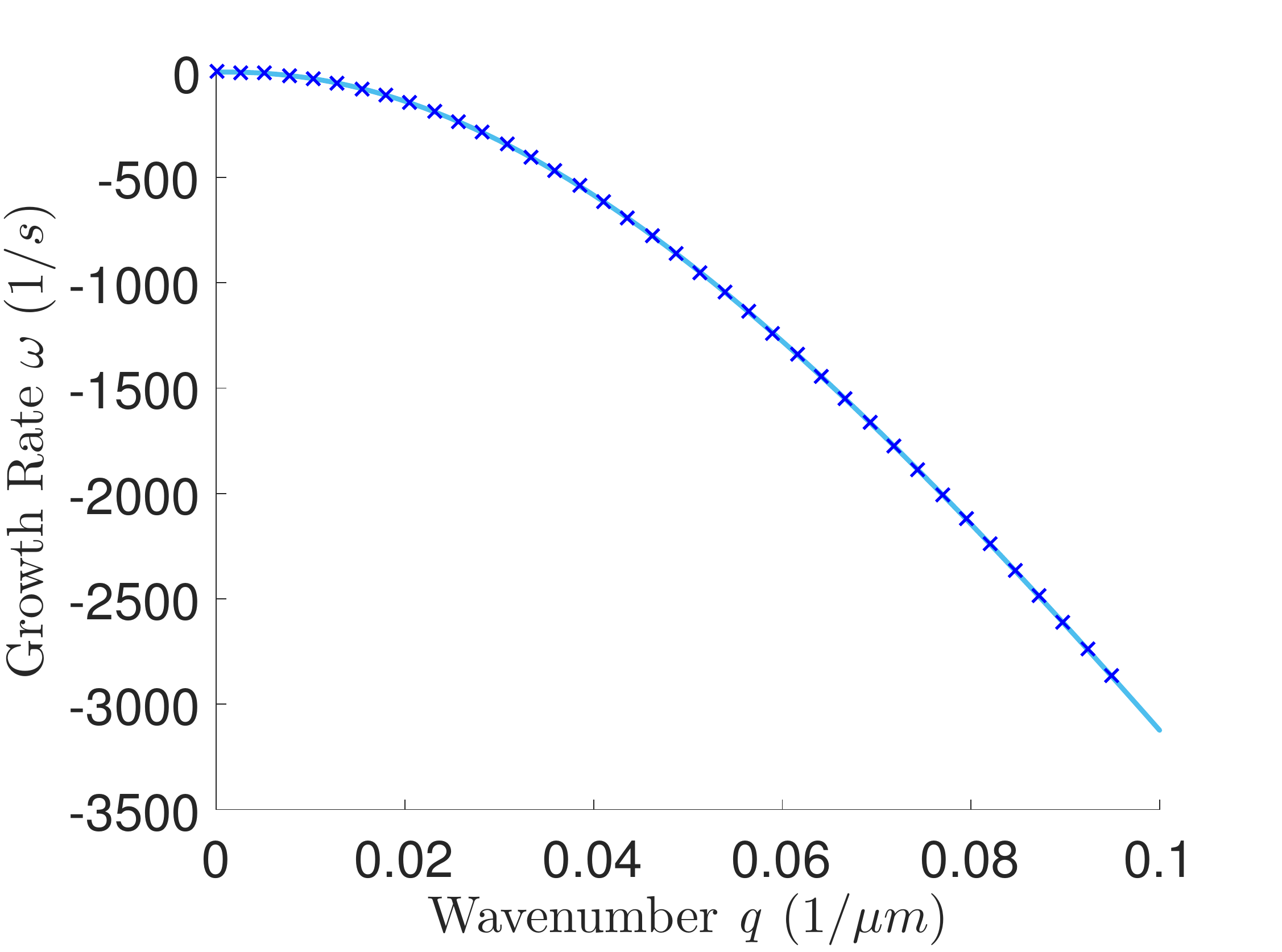} 

   \caption{
   	   	Stability diagram for the moving case ($v_0 = \sqrt{a/b}$) showing the growth rate versus the wavenumber of the perturbation. Results are numerical (crosses) and plotted alongside the solution obtained after ignoring inertial terms (curve).  	   	
   	   	   	   	All parameter values are the same as those in FIG. \ref{fig:Stationary}.      	   	   	
         	   		In addition, we set  $\lambda = 1$ for simplicity.}
   \label{fig:MovingCompare}
  \end{figure}

{\it Moving case.}~~In the case $v_0 = \sqrt{a/b} $ the roots of \eq \eqref{eqn:Quartic} can not be written explicitly in a useful format, hence we present only numerical results.  As all numerical solutions obtained found $\omega$ to be real, we assume that this is true generally. As for the stationary case, $\omega$ has multiple values for each $q$. There are also degenerate cases which were found to be removable (see SM II). Fig. \ref{fig:MovingCompare} displays the highest roots found for the tested values of $q$, plotted alongside the original results obtained using the analysis of \cite{Zimmermann2014}. They align very well, confirming that the inertial terms have little effect on the dominant behaviour of the solution in the moving case, and it is always stable.

{\it Summary \& Discussion.}~~We have performed a linear stability analysis on a strip of incompressible active fluid governed by the hydrodynamic equation of motion in \eq \eqref{eqn:TonerTu} about a stationary system and a constant uniform flow. The fluid density was assumed to be constant and uniform. We found that from a stationary position, the interface can be unstable, subject to the criterion in \eq \eqref{eqn:InstabilityCondition}, and identified that it describes a balance between the viscosity and active driving force. However, the instability that occurs is of the order of the system size. The moving flow, with velocity $\sqrt{a/b}$, is always linearly stable. Our results are qualitatively different than those obtained in \cite{Zimmermann2014}, due to the inclusion of the inertial terms in our analysis.

In the context of wound healing, the lack of a maximal mode would suggest that our incompressible active fluid model is insufficient to produce the fingering behaviour exhibited in wound healing experiments.	Furthermore, using physiological relevant parameters (see caption of \fig \ref{fig:Stationary}), the minimal $a$ required for instability is $2.5 \text{ Pa s }\mu\text{m}^{-2}$ according to \eq \eqref{eqn:InstabilityCondition}. However, in a typical experiment the time scale is of the order of hours.
If we assume that it takes one hour to achieve a steady state, we would expect dimensionally for $\rho/a$ to be about one hour. Using the same value of $\rho$, the density of water, $a$ would therefore be of the order of $2.8 \times 10^{-13} \text{ Pa s }\mu\text{m}^{-2}$. This is many orders of magnitude smaller than the minimum required for instability, explaining why instabilities of the order of the system size are not observed in experiments. Note that the actual density of the cell layer never needs to be incorporated into the analysis in \cite{Zimmermann2014} since the L.H.S. of \eq \eqref{eqn:TonerTu} is set to zero.

Our work thus strongly indicates that compressibility of the tissue is critical for fingering instability. Indeed, examining wound healing assays reveals a stark difference between the typical diameter of a cell in the bulk of the tissue, and that of a cell in the finger-like protrusions, with the latter being significantly larger \cite{Petitjean2010}. Even without an interface, the velocities of cells within a confluent layer have been shown to vary with cell density \cite{Angelini2011}. All of these suggest variations in the cell density should not be ignored.

Our work has relevance beyond tissue regeneration. Since we have shown that the boundary perpendicular to the moving direction of 2D incompressible active fluids is stable, our work suggests that the recent predictions on the scaling behavior of incompressible active fluids in the moving phase \cite{chen16} may be studied on an open system, thus potentially facilitating the experimental verification of the theory.

\begin{acknowledgments}
CFL thanks Bertrand Lacroix-\`{a}-chez-toine for his early involvement with the project and  Amitabha Nandi for discussion.
\end{acknowledgments}

\bibliographystyle{ieeetr}

\onecolumngrid
\newpage

\section*{Supplemental Material}

\section{Deriving $f(\omega,q)$} \label{SM:FDerivation}

The boundary conditions provide a homogeneous system of linear equations in $A_j$ ($j=1,...,4$) and $h_0$. After factoring some arbitrary non-zero terms, such as $\ii$ and $\exp(\ii q + \omega t)$, out of the equations, the system can be expressed as 
\beq
M \mathbf{v} = \mathbf{0}
\eeq

where
\begin{gather} \label{eqn:Matrixd}
	M = \left( \begin{array}{ccccc}
		\ee^{- L r_1} & \ee^{- L r_2} & \ee^{-L r_3} & \ee^{-L r_4} & 0 \\
		\frac{r_1}{q}\ee^{-L r_1} & \frac{r_2}{q}\ee^{-L r_2} & \frac{r_3}{q}\ee^{-L r_3} & \frac{r_4}{q}\ee^{-L r_4} & 0 \\
		1 & 1 & 1 & 1 & -\omega \\
		\frac{r_1^2+q^2}{q} & \frac{r_2^2+q^2}{q}  & \frac{r_3^2+q^2}{q}  & \frac{r_4^2+q^2}{q}  & 0 \\```
		c_1 & c_2 & c_3 & c_4 & \gamma q^2 \end{array} \right) \ ,
\end{gather}

\begin{gather*} 
	c_j =  \frac{ \mu r_j (3 q^2 - r_j^2) + (  \rho \omega - (a - b v_0^2) ) r_j + \rho v_0 (1- \lambda) r_j^2 }{q^2} \ ,
\end{gather*}
and $\mathbf{v} = (A_1,A_2,A_3,A_4,h_0)^T$.
For the solution to be non-trivial we require that $M$ has vanishing determinant. Because the $r_j$'s are the roots of \eq (3), and thus determined by $\omega$ and $q$, the determinant $f(\omega,q) = \det M$ is a function of $\omega$ and $q$. 

\section{Degenerate Solutions} \label{SM:Degenerate}

Some roots of $f(\omega,q)$ are not solutions to our system. If any of the columns of the matrix $M$ are identical, then the determinant is trivially zero. This occurs if any of the roots $r_j$ are repeated, that is, if the quartic (\eq (3)) is degenerate. This situation arises along a small number of curves, $\omega_i=\omega_i(q)$. In these cases the form of the general solution is no longer given by \eq (4), and $M$, as given in \eq \eqref{eqn:Matrixd}, is not valid. Therefore $f(\omega,q)$ cannot be used to determine whether or not $\omega_i(q)$ is a solution; $f$ is trivially zero. The analysis must be repeated in these cases, using the correct general solution and deriving a new $f_i(\omega_i(q),q)$. This new function depends only on $q$, so the solutions we seek are pairs $(\omega_i(q^*),q^*)$ where $f_i(\omega_i(q^*),q^*)=0$. 
Here we derive the conditions for degeneracy and analyse them in turn.

For $v_0 = 0 $, \eq (3) reduces to
\beq
\mu r^4 - (\rho \omega + 2 \mu q^2 -a)r^2 + q^2(\mu q^2 + \rho \omega - a) =0
\eeq
hence
\beq
r^2 = q^2 + \frac{(\rho \omega -a) \pm (\rho \omega  -a)}{ 2\mu}
\ .
\eeq
This is degenerate in two cases
\beq
\omega_1 = \frac{a}{\rho}  \sep\omega_2 = \frac{a}{\rho} - \frac{\mu}{\rho}q^2
\ .
\eeq
In the moving case, $v_0 = \sqrt{a/b}$, the quartic equation (3) cannot be solved in a useful format for general parameters, hence for simplicity we restrict our attention to the special case $\lambda = 1$. Therefore \eq (3) becomes
\beq
\mu r^4 - (\rho \omega + 2 \mu q^2)r^2 + q^2(\mu q^2 + \rho \omega + 2a) = 0
\ ,
\eeq
so
\beq
r^2 = \frac{1}{2\mu}(\rho \omega + 2 \mu q^2 \pm \sqrt{ \rho^2 \omega^2 - 8 \mu a q^2} ).
\eeq
Roots are repeated in the cases
\beq
\omega_3 = - \frac{2a}{\rho} - \frac{\mu}{\rho}q^2 \sep \omega_{\pm} = \pm \frac{\sqrt{8 \mu a}}{\rho} q
\ .
\eeq
The former provides only negative values of $\omega$, which are stable and thus do not require further analysis. The remaining cases must be treated separately and a corresponding $M$ must be derived, which must once again have zero determinant.

In what follows, let $u_x = \ee^{\ii qy + \omega t} \tilde{u}_x(x-v_0t)$, and similarly for $u_y$ and $p$.

\subsection{\textit{Stationary Case: $\omega_1 = a/\rho$}}

The general solution is
\beq
\tilde{u}_x = (A_1 + A_2x)\ee^{qx} + (A_3 + A_4 x) \ee^{-qx} .
\eeq
The corresponding $\tilde{u}_y$ and $\tilde{p}$ are obtained from \eq (1) and given by
\begin{align}
	\tilde{u}_y =\frac{\ii}{q}& \big[ ( A_1 q + A_2(1+qx) )\ee^{qx}  \\
	&-  (q A_3 + A_4 (-1+ qx)) \ee^{-qx} \big]  ,         \nonumber \\
	\tilde{p}=\frac{1}{q^2}& \Big\{    A_1 q(a- \omega \rho)\ee^{qx}  \nonumber \\
	&+ A_2\left[(a- \omega \rho)(1+qx)+2 q^2 \right]\ee^{qx}   \\
	&-  A_3q(a- \omega \rho)\ee^{-qx}  \nonumber  \\  
	&+  A_4  \left[(a- \omega \rho)(1- qx) + 2 q^2 \right]\ee^{-qx} \Big\} . \nonumber
\end{align}
As before the boundary conditions provide a linear system of equations in $A_j$ and $h_0$. The determinant of the matrix of coefficients must be zero, which reduces to
\beqn \nonumber
0 &= & 2 + 4 (Lq)^2 + 2 \cosh{2Lq} \\
&& + \frac{\rho \gamma}{a \mu} \left( q \sinh(2Lq) - 2(Lq)^2 \right)\ .
\eeqn
This equation has no solutions for any real $q$ regardless of the parameter values, hence the root $\omega_1 = a/\rho$ can be disregarded.

\subsection{\textit{Stationary Case: $\omega_2 = a/\rho - (\mu/\rho)q^2$}}

The general solution is
\beq
\tilde{u}_x = A_1 \ee^{qx} + A_2\ee^{-qx} + A_3 x + A_4 .
\eeq
The corresponding det$M =0$ reduces to
\beqn \nonumber
0&=& \frac{\mu}{\rho} (\mu q^2 - a)(4 - 5 \cosh(Lq) + Lq \sinh(Lq))  \\
&&\label{eqn:f2}
+ \gamma q (L q \cosh(Lq) - \sinh(Lq)) \ . 
\eeqn
It is not obvious how many solutions there are to this equation, if any at all; it depends heavily on the parameter values. For the values used in \fig 2 there are exactly two solutions for which $\omega_2>0$. These match the points where the curve $\omega_2 = a/\rho - (\mu/\rho)q^2$ intersects the curves obtained from the non-degenerate cases. Assuming this result applies for all parameter sets, we could garnish information about the full solution from the number of valid points on the degenerate curve $\omega_2(q)$. For example, if there are no such solutions to \eq \eqref{eqn:f2}, then we can conclude that all of the unstable solution curves are bounded below $\omega = a/\rho - (\mu/\rho)q^2$. Otherwise we would have discontinuities in our full solution.

\subsection{\textit{Moving Case: $\omega_+ = + \sqrt{8 \mu a} q/\rho$}}

The general solution has the form
\begin{align}
	\tilde{u}_x =& \quad (A_1 + A_2(x-v_0t))\ee^{r(x-v_0t)} \\ 
	&+(A_3 + A_4(x-v_0t)) \ee^{-r(x-v_0t)} \nonumber
\end{align}
where $r$ is the positive solution to
\beq \label{eqn:r2}
r^2 = q^2 + \sqrt{2 a / \mu }\text{ }q .
\eeq
It is not apparent what the corresponding $\tilde{u}_y$ and $\tilde{p}$ are, so we assume:
\begin{align}
	\tilde{u}_y =& (\hat{B}_1 + \hat{B}_2)\ee^{r\xi} + (\hat{B}_3 +\hat{B}_4) \ee^{-r\xi} , \\
	\tilde{p}   =& (\hat{C}_1 + \hat{C}_2)\ee^{r\xi} + (\hat{C}_3 + \hat{C}_4) \ee^{-r\xi},
\end{align}
where the $\hat{B}_j$ and $\hat{C}_j$ are all {\it functions} of $\xi = x-v_0 t$, and relate directly to $A_j$, for each $j$ respectively. Using the incompressibility condition we obtain
\begin{align}
	\tilde{B}_1 = \frac{\ii r}{q} A_1   \qquad,\qquad  \tilde{B}_2 = \frac{ \ii (1 + r \xi)}{q} A_2   \qquad,\qquad
	\tilde{B}_3 = -\frac{ \ii r}{q} A_3   \qquad,\qquad \tilde{B}_4 = \frac{\ii(-1 + r \xi)}{q} A_4 \ .
\end{align}
By substituting $\tilde{B}_j$ into the momentum equation for $u_y$, we can obtain
\begin{eqnarray}
\tilde{C}_1 &=& \qquad - \frac{r (\mu (q^2 - r^2) + \rho \omega )}{q^2} A_1 \ , \\
\tilde{C}_2 &=&- \frac{ (\mu (q^2 - r^2) + \rho \omega) (1+r \xi) -2\mu r^2 }{q^2} A_2 \ , \\
\tilde{C}_3 &=& \qquad \frac{r (\mu (q^2 - r^2) + \rho \omega )}{q^2} A_3  \ , \\
\tilde{C}_4 &=&  - \frac{ (\mu (q^2 - r^2) + \rho \omega) (1 - r \xi) -2\mu r^2 }{q^2} A_4 \ .
\end{eqnarray}
Applying the boundary conditions and factoring out constants, we get a matrix system $ M \mathbf{v}=0$ as before, this time:
{\small \begin{gather} \label{eqn:MatrixDegenAlt}
		M= \left( \begin{array}{ccccc}
			\ee^{- L r} & -L \ee^{- L r} & \ee^{L r} & -L\ee^{L r} & 0 \\
			r \ee^{-L r} & (Lr-1)\ee^{-L r} & -r \ee^{L r} & (Lr+1)\ee^{L r} & 0 \\
			-1 & 0 & -1 & 0 & \omega \\
			r^2+q^2 & 2r  & r^2+q^2  & -2r  & 0 \\
			c_1 & c_2 & -c_1 & c_2 & \gamma q^2 \end{array} \right)
\end{gather} }
where
\beqn
c_1 &=&  r (\mu  (3 q^2 - r^2) +   \rho \omega ) \ , \\
c_2 &=&   3 \mu  ( q^2 - r^2) +   \rho \omega \  .
\eeqn
Once again the determinant of this matrix must be zero, explicitly
\begin{align}
	0=& 2 \gamma q^4 r ( 2 \sinh{2Lr} - 4 L r) \nonumber \\
	&+ \mu \omega \left\{  r^4(2 \cosh{2Lr} + 14 - 4L^2r^2 ) \right. \nonumber\\
	&  \qquad+ q^2 \left[ r^2 (8L^2r^2 + 12 \cosh{2Lr} -12) \right.\nonumber \\
	&  \qquad \qquad	\left. \left. + q^2 ( 12 L^2r^2  + 6 - 6 \cosh{2Lr}) \right] \right\} \\
	& + \omega^2 \rho\left\{ r^2(2 \cosh{2Lr} -2 +4L^2r^2) \right. \nonumber \\
	&  \qquad  \left. - q^2(2 \cosh{2Lr} -2 - 4L^2r^2) \right\} . \nonumber
\end{align}
By examining each of these lines we can deduce that there are no solutions. Recall that all of our parameters are positive, including $r,q$ and $\omega$. By inspection the first two lines are strictly positive individually.
The square brackets straddling lines 3 and 4 are also strictly positive, seen using the fact that $r^2>q^2$ from \eq \eqref{eqn:r2}.
Similarly the curly brackets straddling lines 5 and 6 are also strictly positive. Hence there is no way that the expression on the right hand side can be zero, so we conclude that the degenerate root $\omega_+ = + \sqrt{8 \mu a} q/\rho$ is not a solution to our system, and can be disregarded from numerical results.

\end{document}